\documentclass[runningheads]{llncs}
\usepackage[T1]{fontenc}
\usepackage{array}
\usepackage{tabularx}
\usepackage{booktabs}
\usepackage{url}
\usepackage{geometry}
\usepackage{algorithm} 
\usepackage{algpseudocode} 
\usepackage{graphicx}
\begin{document}
\title{A Scientific Data Integrity system based on Blockchain}
%
\author{Gian Sebastian Mier Bello\inst{1}\orcidID{0009-0003-9276-1017} \and
A Martínez-Méndez\inst{1,2}\orcidID{0000-0002-1559-9015} \and
Carlos J. Barrios H.\inst{1,3,4}\orcidID{0000-0002-3227-8651}  \and
Robinson Rivas\inst{5}\orcidID{0000-0001-7900-0764} \and
Luis ~A.~Núñez\inst{1,2}\orcidID{0000-0003-4575-5899}
}
\authorrunning{G. Mier et al.}
%
\institute{SC3UIS-CAGE, Universidad Industrial de Santander, Bucaramanga, Colombia 
\and
GIRG, Universidad Industrial de Santander, Bucaramanga, Colombia \and
LIG/INRIA DataMove, Grenoble, France\and
CITI/INRIA SiNDY, Lyon, France \and
Universidad Central de Venezuela, Caracas, Venezuela\\
\email{gian2210073@correo.uis.edu.co}, \email{alexander2198160@correo.uis.edu.co }, \email{cbarrios@uis.edu.co}, \email{robinson.rivas@ciens.ucv.ve}, \email{lnunez@uis.edu.co}
}
\maketitle              
\begin{abstract}
In most High Performance Computing (HPC) projects nowadays, there is a lot of data obtained from different sources, depending on the project's objectives. Some of that data is very huge in terms of size, so copying such data sometimes is an unrealistic goal. On the other hand, science requires data used for different purposes to remain unaltered, so different groups of researchers can reproduce results, discuss theories, and validate each other.  In this paper, we present a novel approach to help research groups to validate data integrity on such distributed repositories using Blockchain. Originally developed for cryptographic currencies, Blockchain has demonstrated a versatile range of uses. Our proposal ensures 1) secure access to data management, 2) easy validation of data integrity, and 3) an easy way to add new records to the dataset with the same robust integrity policy. A prototype was developed and tested using a subset of a public dataset from a real scientific collaboration, the Latin American Giant Observatory (LAGO) Project.

\end{abstract}
%
%
%

\section{Introduction}
Almost all pieces of data published on the Internet today are at risk of being manipulated or viewed without prior authorization; data can be manipulated at different layers of the network. In this context, Blockchain appeared, which was initially conceived as a P2P transaction system, but thanks to its properties such as decentralization, immutability, and encryption, it has proven to be effective in ensuring the integrity and traceability of digital records. The most recognized application of this technology is Bitcoin. As per the observations made by Hirsh \& Alman \cite{ref_bookHirshAlman}, Bitcoin facilitates secure monetary transactions without the need for financial intermediaries. Nevertheless, they assert that the primary objective of this technology is to establish an immutable record of ownership and provenance. This characteristic is essential for ensuring the integrity of scientific data and, ultimately, the advancement of scientific research.

HPC has become essential in recent times for the advancement of new scientific knowledge and, consequently, for scientific research across various disciplines, as articulated by Silva et al. \cite{ref_Silva}. In consideration of modern computational capabilities, it is now possible to execute complex simulations with a degree of realism and precision unprecedented. Nowadays, numerous applications requires substantial computational resources encompass seismic analysis related to natural disasters, oil and gas exploration, climate and flood forecasting, computational fluid dynamics (CFD), Deoxyribonucleic Acid (DNA) sequencing, simulations of electromagnetic equations, analyses in the energy sector, materials science, and other related domains \cite{ref_Mantripragada}.

The applications and resources are optimized and calibrated to achieve superior performance in their execution, resulting in the generation of substantial volumes of data. Thus, it is imperative to manage these resources appropriately to facilitate research. A pertinent example of such projects is the Latin America Giant Observatory (LAGO) project \cite{lago}. Within this initiative, data are derived from a network of radio telescopes and subsequently processed utilizing high-performance computing (HPC) facilities in Latin American countries. It is of utmost importance that, for scientific purposes, the data remain unaltered and accessible, enabling various research groups to collaborate reliably and discuss findings. 
In instances such as the LAGO project, the substantial size of the data poses significant challenges in maintaining trustworthy copies of the entire datasets. The proposed system addresses this by adopting an off-chain storage model; only the cryptographic hashes and metadata of the scientific data are stored on the blockchain, while the large datasets themselves remain in their original distributed repositories. This approach ensures data integrity without overloading the blockchain, which is not designed to store large files, thereby avoiding constraints on block size. Therefore, in this work, we propose a prototype to support the integrity of the measurement and simulation data of the LAGO project.

Combining blockchain technology with high-performance computing (HPC) is an emerging field that could significantly change how computational resources are utilized and shared. This novel approach presents challenges in management and usage, including scalability, consensus mechanisms, blockchain architectures, and widespread adoption in large-scale scientific infrastructures. However, the benefits in trust, security, cost efficiency, and performance in heterogeneous systems (including levels in multiscale architectures such as Edge and Fog Computing) as well as decentralized advanced computing capabilities motivate the research, development, deployment, and implementation of this integration.

In this context, the proposed blockchain-based prototype contributes by addressing key challenges in data integrity and access control within distributed HPC environments. This can be achieved by leveraging Hyperledger Fabric’s permissioned architecture, which restricts participation to authenticated entities and users  using cryptographic identities (X.509 certificates) with \textbf{MSP}. Access policies are enforced through chaincode logic, allowing fine-grained control over who can register, query, or update metadata linked to scientific datasets. Regarding traceability, each transaction—including the registration of raw data, simulation inputs, and processed outputs—is immutably recorded in the ledger, providing a verifiable and timestamped audit trail of the entire data lifecycle. This enables researchers to reconstruct workflows, validate results, and attribute contributions across distributed HPC infrastructures. For tamper-evidence, the use of cryptographic hashing ensures that any alteration to input/output files or metadata would result in a mismatch with the stored hash, and the content block hash in the chain too, as a result signaling potential integrity breaches. This technique allows to ensure data has not being manipulated nor changed since the use by research groups. However, it doesn't prevent the data manipulation itself: only serves as a warrantee of data pureness if any other group wants to check for data integrity or compare sources.

This paper is structured as follows: Section \ref{background} provides background on the LAGO project, which offers a public dataset emphasizing transparency, low cost, and heterogeneity integration; and the state of the art and key features revised to implement our proposal. Section \ref{EE} discusses the execution environment, considering the analysis to choose the optimal deployment strategies, designing the prototype, demonstrating the proof of concept, and validating and fine-tuning the deployed prototype. In section \ref{results}, we present the most significant results, followed by a discussion in section \ref{discussion}. Finally, we conclude with our findings and suggestions for further work in section 
\ref{Conclusion}.

\section{Background}
\label{background}
In this section, we present the basis of our work: the LAGO initiative and their need for data, and some state-of-the-art on Blockchain technology and its relation with Scientific Data Management

\subsection{LAGO}
The Latin American Giant Observatory (LAGO) constitutes a sophisticated astroparticle observatory established to examine space weather phenomena, extreme astrophysical events, and atmospheric radiation at ground level. It functions through a distributed network of Water Cherenkov Detectors (WCD)  strategically deployed at thirty-one sites, encompassing a broad spectrum of latitudes, from Mexico to Antarctica, and altitudes ranging from sea level to 5000 meters covering a wide range of geomagnetic rigidity cut-offs and atmospheric absortion/reaction levels. The LAGO WCD is simple and robust, and incorporates several integrated devices to allow time synchronization, autonomous operation, on-board data analysis, as well as remote control and automated data transfer \cite{Sidelnik_2017} \cite{SIDELNIK2020161962}. These detectors collect substantial volumes of cosmic ray measurements across different environmental and geomagnetic conditions, yielding an extensive and unique scientific dataset.

\begin{figure}[!htbp]
\centering
\includegraphics[width=0.5\textwidth]{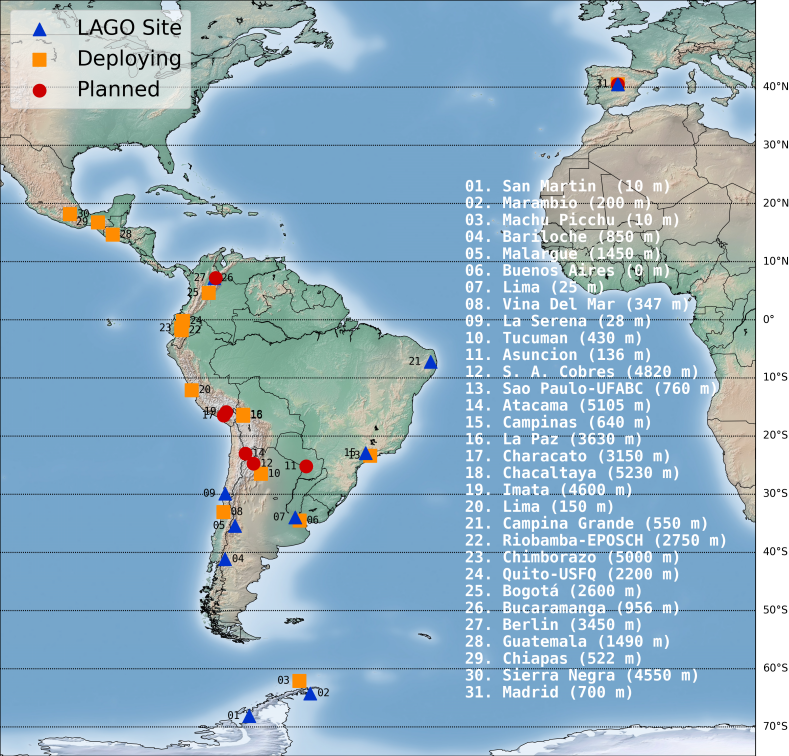}
\caption{Geographical distribution of LAGO Project WCD detectors. Taken from \cite{ref_Rubio}} \label{fig1}
\end{figure}

Figure \ref{fig1} shows the different sites where detectors are placed, indicated by blue triangles. The orange squares represents the deploying sites where computing capabilities and data storage are available, and the red circles indicate the planned sites.


LAGO is coordinated by the LAGO Collaboration, which comprises over 90 researchers from 25 institutions across Latin America and Spain. In addition to collecting experimental data, the collaboration also produces simulation data using high-performance computational (HPC) frameworks. Both experimental and simulated datasets are organized into hierarchical layers according to their level of processing and intended scientific use, with higher layers significantly reducing the data volume.

One of LAGO’s primary objectives\cite {ref_LAGOdmp} is to preserve both raw and processed data for potential reprocessing, particularly in the event of software errors, changes in analysis pipelines, or methodological updates. This is crucial given the computational cost, energy consumption, and complexity involved in regenerating degraded or lost data. Furthermore, as HPC environments are heavily used in LAGO for executing simulation and data analysis workflows, ensuring the integrity and verifiability of their outputs is fundamental for scientific reproducibility. In this context, this work proposes a blockchain-based solution designed to integrate with HPC workflows and support the secure registration of data products. By enabling tamper-proof, traceable, and verifiable records of datasets across the collaboration, the system contributes to more reliable data management, facilitating cross-institutional trust and reducing unnecessary reprocessing efforts.

\subsection{State of the art}

Some works related to the use of blockchain in HPC to ensure data integrity and promote scientific collaborative environments were identified. However, no approach was found that simultaneously encompasses these issues.

A first relevant work is “A Blockchain‑Based Architecture for Trust in Collaborative Scientific Experimentation \cite{ref_Coelho} In this study, a cloud-hosted blockchain network with geographically distributed nodes was implemented to support and build trust in scientific collaborative environments. The solution leverages the immutability of records in blockchain to ensure data provenance and uses a decentralized application that collects key data based on a provenance model. Their case study focused on coronavirus genomic data, where access management was performed using certificates that guarantee the identity of contributors. Although this paper shares our vision regarding data integrity, we focus on supporting a high computation environment in different geographic zones.

Another relevant work is “SciLedger: A Blockchain-based Scientific Workflow Provenance and Data Sharing Platform” \cite{ref_Hoopes}. This solution implements a blockchain to support scientific workflows, with features such as handling multiple streams and an invalidation mechanism. It uses a Merkle tree to verify the provenance of data and allows public access to the stored data stream.
While sharing the vision of ensuring data integrity, our approach differs by prioritizing the data flow support in every layer.

Finally, in “Earth Observation Data Provenance: A Blockchain-Based Solution” \cite{ref_Zhang}, the traceability and sharing of data related to observations of the Earth's surface and atmosphere are addressed. Due to the volume of data in petabytes, this work proposes an off-chain storage model, using the blockchain to store key information using hashes. In addition, it uses a voting-based consensus algorithm, optimizing performance and energy efficiency.  In this case, many similarities are shared in terms of how to ensure traceability, store, and guarantee the  integrity of the data using a decentralized application. Taken together, these works demonstrate the potential of blockchain technology to strengthen collaborative science. They provided the foundation for this proposal; however, none of them focused on verifying the integrity of the data.

\section{Execution environment}
\label{EE}
This work was developed based on "7 phases of Blockchain Implementation" \cite{ref_CompTIA}. The following steps were: review of the state of the art, analysis and selection, prototype design, proof of concept, and validation and adjustment.

\subsection{Prototype Architecture}

After a detailed study of the LAGO project and its data flow in the collaboration, the options were limited to network types with permissions. Table \ref{tab:blockchain_comparison} provides a comparison of the leading permissioned blockchain platforms considered for this work.

\begin{table}[!htbp]
\centering
\scriptsize
\begin{tabularx}{\textwidth}{|>{\bfseries}l|X|X|X|X|}
\hline
\textbf{Feature} & \textbf{Hyperledger  Fabric \cite{ref_Raj}} & \textbf{Ripple \cite{ref_Raj}} & \textbf{Corda \cite{ref_Raj}} & \textbf{Quorum \cite{ref_Raj}} \\
\hline
Privacy & Private channels for confidential transactions & Trust network based on credit relationships & Transactions visible only to involved parties & Private transactions via nodes and Tessera \\
\hline
Consensus & RAFT & Validator nodes & Validation by notaries & RAFT, IBFT \\
\hline
Smart contracts & Modular support with Chaincode & Not applicable & JVM with legal regulations & Compatible with Solidity and EVM \\
\hline
Automation & Chaincode for business rules & Limited to settlements & Based on legal contracts & Compatible with Ethereum tools \\
\hline
Main applications & General business sectors & Finance, remittances, and global payments & Financial institutions and legal sectors & Finance, logistics, and industries using Ethereum \\
\hline
\end{tabularx}
\caption{Comparison of blockchain platforms}
\label{tab:blockchain_comparison}
\end{table}

Upon analyzing the other alternatives in the table—Ripple, Corda, and Quorum—it was observed that they all focus on the financial aspect. Therefore, Hyperledger Fabric 2.5 \cite{ref_HF} is the most suitable option for the LAGO project, as its network with permissions structure ensures that only authorized participants can access the data. It also supports private channels and smart contracts, improving confidentiality and operational efficiency. Being open source allows for customization without licensing costs and makes it easy to adapt to the project's needs. Furthermore, its consensus mechanism, RAFT, operates by choosing a leader to validate transactions even in the event of communication failures. This reduces energy and hardware costs, making its implementation more sustainable in the long term.

\begin{figure} [h]
        \centering
        \includegraphics[width=0.8\linewidth]{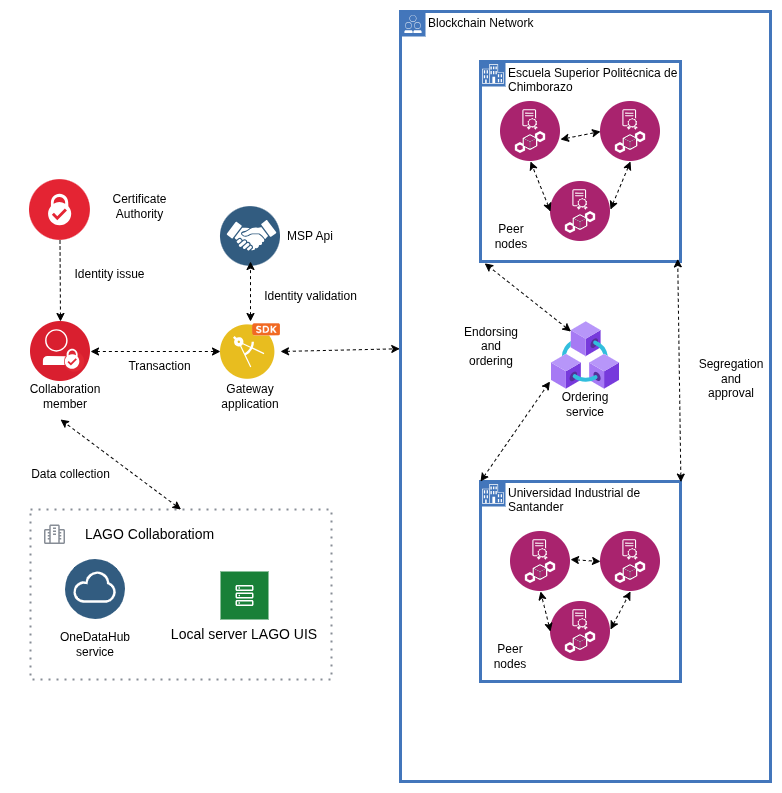} 
        \caption{Blockchain prototype architecture}
        \label{fig:ModelArchitecture}
\end{figure}

The blockchain prototype shown in Fig. \ref{fig:ModelArchitecture} uses Hyperledger Fabric as its base platform. Two member organizations of the LAGO Collaboration were taken into account: \textbf{Universidad Industrial de Santander} and \textbf{Escuela Superior Politécnica de Chimborazo}, since both had the necessary data for the test. To support the RAFT protocol based on Crash Fault Tolerance (CFT), an odd number of nodes per organization was chosen.  The description of the components and the transaction flow is detailed below.

\begin{enumerate}
\item  \emph{Certificate Authority} is responsible for issuing MSP (Membership Service Provider) based digital identities to network participants. These identities are encapsulated in X.509 certificates, which are required to interact with the network.

\item \emph{Chaincode} The chaincode defines the business logic and rules of the blockchain network, managing key-value pairs that represent the initial state and transactions of the system. Unlike traditional smart contracts, chaincode operates on authorized private networks within nodes rather than on decentralized virtual machines, under specific endorsement policies that guarantee security and control.
\item \emph{Gateway application (SDK)} is the bridge between users (collaboration members) and the blockchain network. It facilitates the submission of transaction proposals and identity validation using the MSP API. Transactions refer to Chaincode functions running on the nodes.

\item \emph{Ordering service} It is responsible for receiving transactions proposed by users, assigning them a chronological order, and distributing the resulting blocks to peer nodes in both organizations.

\item \emph{Peer nodes} Each organization has three peer nodes that validate the transactions according to the endorsement policies and store a copy of the updated blockchain. This ensures the decentralization of the recorded data.

\item \emph{LAGO Collaboration} represents the integration of the data collection systems, such as the OneDataHUB Service and the local LAGO UIS server, which receives the measurement data (L0), with the blockchain network. The measurement data and transactions by the collaboration members are processed and stored by transactions.
\end{enumerate}

\subsection{Data scheme}


The proposed data schema defines the generic structure of information recorded in the Hyperledger Fabric blockchain network to store scientific simulations and measurements of the LAGO project. Each record has an \textbf{Id}, which acts as a unique identifier and follows the DMP convention \cite{ref_LAGOdmp}; a \textbf{Type} field, which indicates whether it is a simulation or measurement and its level (L0, L1, S0, S1, etc. ); \textbf{Metadata}, corresponding to the metadata of the raw data; \textbf{Raw data}, with the hash and location of the data obtained; \textbf{Input data} and \textbf{Input metadata}, which store the input data and metadata used in simulations; \textbf{Output data} and \textbf{Output metadata}, which record the processed results and their respective metadata. In addition, additional fields are included such as \textbf{Site name} (name of the site where data was generated), \textbf{Collaborator name} (name of the responsible collaborator), \textbf{ORC id} (ORCID identifier of the collaborator) and \textbf{Access url}, which provides the external address where the original file or related information can be consulted or accessed. This structure ensures that all elements of the scientific workflow are clearly defined, verifiable and accessible.


\subsection{Materials}

For the implementation of the prototype in Hyperledger Fabric we have a x86\_64  GNU/Linux Debian 6.1.128-1 machine with 7.51 GB of RAM and 8 cores, along with the tools: \textbf{Docker} and \textbf{Docker Compose}. \footnote{ Docker and Docker Compose are tools to manage containers: https://www.docker.com/ } essential for running different network services such as containers with the \textbf{images} of certificate authorities (fabric-ca), peers, noSQL CouchDB databases (couchdb) and orderers.

Digital identities were configured using X.509 certificates to ensure controlled access on the network. An inter-organizational communication channel was established, and the `ScientificDataCollectContract` chaincode was deployed to manage scientific records with role-based access control.

\begin{algorithm}
\scriptsize
\caption{Verify Block Integrity}
\begin{algorithmic}[1]
\Procedure{IntegrityVerification}{contentHash, dataHashString, block, i}
    \If{contentHash $\neq$ dataHashString}
        \State $blockDataList \gets$ Get the list of transactions from the block.
        \If{$blockDataList$ is not $empty$}
            \For{each $bl$ in $blockDataList$}
                \If{$bl$ is not $empty$}
                    \State $signature \gets$ Extract the transaction signature.
                    \State $envPayload \gets$ Get the transaction payload.
                    \State $headerPayload \gets$ Extract the transaction header.
                    \State $txId \gets$ Get the transaction identifier. 
                    \State $creatorBytes \gets$ Extract the transaction creator's identity. 
                    \State $verifier \gets$ Prepare the signature verification using the SHA-256 algorithm. 
                    \State Validate the creator’s signature with the transaction data.
                    \State $isValid \gets$ Sign $envPayload$ with $creatorBytes$ and compare with $signature$ 
                    \If{$isValid = false$}
                        \State Throw an error indicating that block $i$ has been tampered with in the transaction with ID $txId$
                    \EndIf
                \Else
                    \State Print No transaction data in block
                \EndIf
            \EndFor
        \EndIf
    \EndIf
\EndProcedure
\end{algorithmic}
\label{BlockIntegrity}
\end{algorithm}

It was identified that if a peer node falls out of sync with the network—due to connectivity loss or other faults—it may locally possess a divergent or outdated version of the ledger. Transactions submitted to such a node risk being rejected by the broader network during the ordering and validation phases because they are based on an inconsistent world state. This behavior is not an oversight by the protocol, but a fundamental security feature to prevent "double-spending" or the introduction of invalid data. To mitigate this, the client application was designed to halt transaction proposals to any peer whose ledger state is detected as corrupt or inconsistent, as detailed in Algorithm \ref{BlockIntegrity}.

This \textbf{gateway} application was developed in \textbf{Node.js} to interact with the network. It verifies the integrity of a block on the blockchain by comparing the stored data with that calculated in real time. First, the hash of the content (contentHash) is compared with the hash recorded in the block header (dataHashString). If they do not match, it is considered that the block might have been tampered with. Then, the transactions within the block are extracted and verified.

For each transaction in the block, if it is not empty, the signature, payload, and header are extracted, along with the identifier (txId) and the identity of the originator. A signature verifier using SHA-256 is prepared and validated if the creator's signature is authentic. If the validation fails, an error is thrown indicating that the transaction with the txId in block i has been tampered with. If there is no transaction data, a message is printed indicating its absence. The implementation repository is available at \footnote{Implementation repository: https://github.com/GSMier/ProyectoLago}

\subsubsection{Data collection}

The process started with the extraction of a subset of data from two main sources. On the one hand, the measurements stored in a local server of the
LAGO project were selected and downloaded. On the other hand, the simulations were obtained from the centralized repository of One Data Hub, a platform for storage and access of scientific data. These two sources provided the base information to be processed.

The extracted data were processed using Python scripts, designed to convertthe measurements and simulations into a standard, lightweight JSON-like structure. During this processing, the input and output files and their respective metadata were analyzed. To ensure data integrity, unique identifiers were generated using the hash function SHA-256, applied to each of the files and their contents to the original format. This approach allows verifying that the data have not been altered during or after
processing.

Finally, the processed data and the scripts used for its transformation were stored in a GitHub repository. This repository, accessible at \footnote{ 
Link to the repository of scripts and processed data https://github.com/GSMier/LAGO-data }, contains the generated JSON, as well as the code used to transform the data. This centralized storage facilitates access to the processed data and allows other collaborators to replicate
the process or adapt it to new needs.

\begin{figure}[ht!]
        \centering
        \includegraphics[width=1\linewidth]{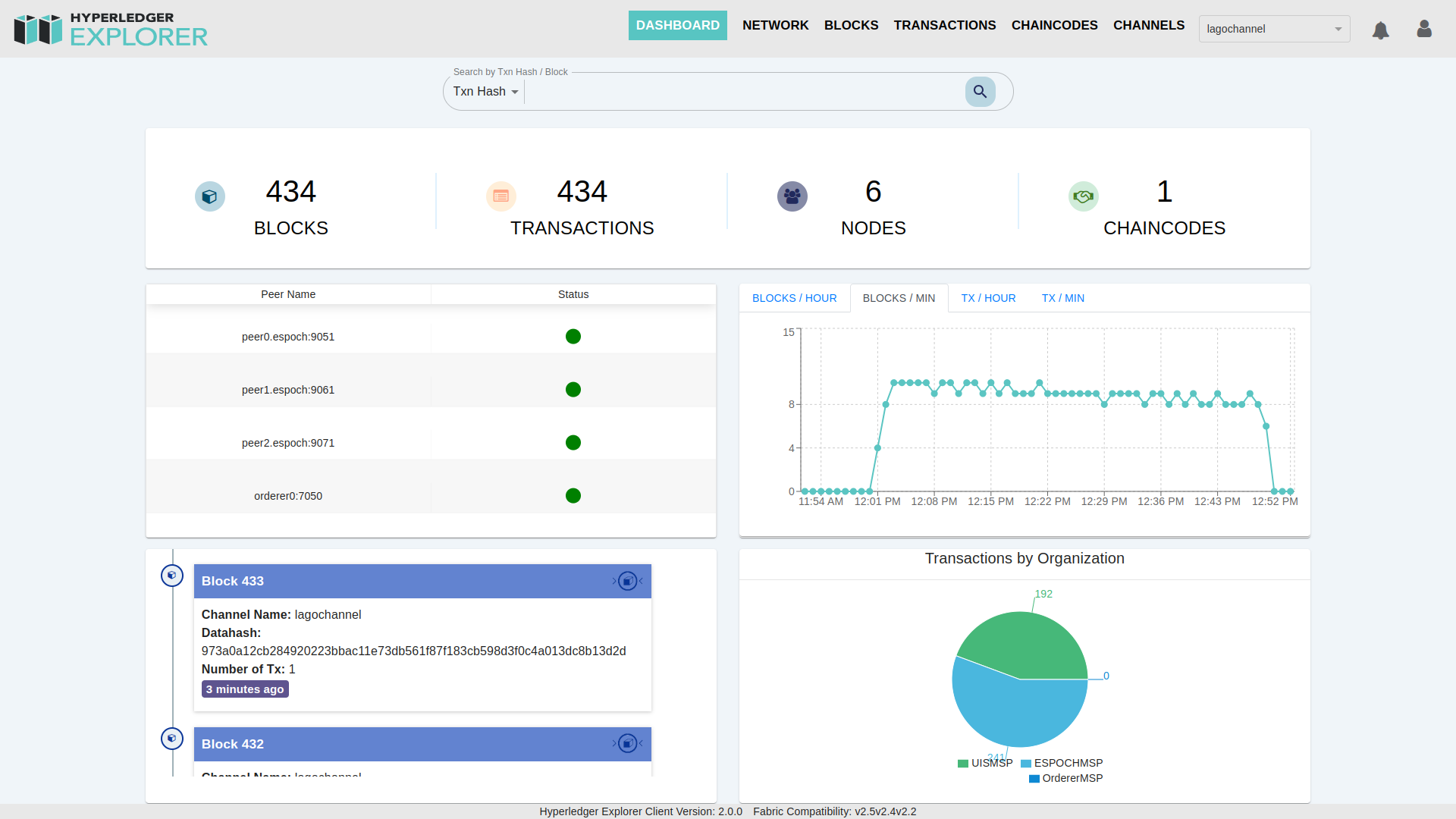} 
        \caption{Hyperledger Explorer Dashboard}
        \label{fig:ExplorerDashboard}
\end{figure}

\begin{figure}[ht!]
        \centering
        \includegraphics[width=0.6\linewidth]{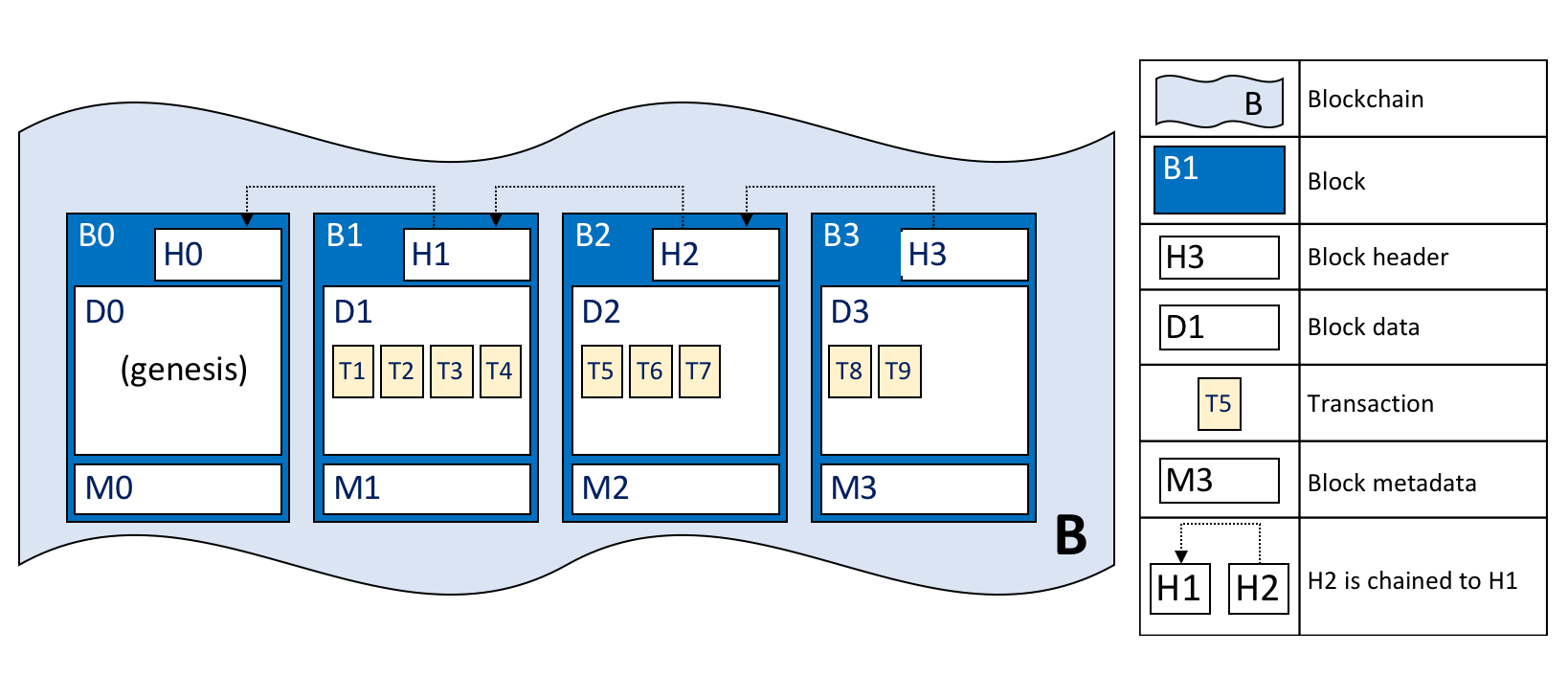} 
        \caption{Hyperledger Fabric blockchain structure  — Taken from \cite{ref_HFDoc}}
        \label{fig:BlockchainStructure}
\end{figure}

The storage in the blockchain was then carried out with previous data.  With the help of \textbf{Hyperledger Explorer}, a web service developed by Hyperledger Fabric \footnote{Hyperledger Explorer repository: https://github.com/hyperledger-labs/blockchain-explorer }, network data such as blocks, nodes, transactions, chaincode, and channels can be visualized. As can be seen in Fig. \ref{fig:ExplorerDashboard}, 434 transactions were performed, of which 430 were invoked by the chaincode "ScientificDataCollection" to store the data, which were distributed in different blocks, since each block accepts transactions in 2-second intervals. With an average of approximately 8 transactions/min, the scientific data of the 2 organizations were recorded.
In Hyperledger Fabric, its structure is more complex, as each transaction includes digital signatures generated with the private keys of its creators. In addition, peer nodes that validate and approve these transactions also add their signatures to the blocks. Fig. \ref{fig:BlockchainStructure} illustrates this structure, where each block is linked to the previous one via cryptographic hashes.

\subsection{Proof of concept}
The blockchain is the component where all transactions recorded on the network are stored. In Hyperledger Fabric, its structure is more complex, as each transaction includes digital signatures generated with the private keys of its creators. In addition, peer nodes that validate and approve these transactions also add their signatures to
the blocks.

In Hyperledger Fabric, the integrity of the data in the blockchain is validated by verifying blocks and transactions. At the block level, the
continuity of the chain is ensured by checking that the hash of the previous block matches the stored hash in the current block, avoiding alterations in the history. It also verifies that the hash of the current block corresponds to its content, ensuring that it has not been modified.
In addition, metadata signatures and configurations are validated to confirm that the blocks have been issued by authorized entities and that the network configuration is consistent and legitimate. For transactions, checks are performed on the digital signatures on the headers and approvals, ensuring that the identities involved belong to authorized entities within the Membership Service Provider (MSP). It is validated that the signatures of the approvers are legitimate and correspond to the content approved in the transaction, ensuring that the consensus has been followed correctly.


In the Hyperledger Fabric blockchain structure, data integrity is enforced by specific cryptographic mechanisms. First, all blocks are cryptographically linked using \textbf{SHA256} hash keys, where every block header contains the hash of the previous block, creating an immutable chain[cite: 164, 165]. Second, every transaction is secured using the \textbf{Elliptic Curve Digital Signature Algorithm (ECDSA)}. Through the \textbf{Membership Service Provider (MSP)}, each participant is issued a public/private key pair. Transactions are signed with the creator's private key, ensuring both authenticity and integrity, as any alteration to the transaction data would invalidate the signature. 
To assess the integrity of the data, the on-chain records of measurements and simulations—such as the data hashes, metadata, and collaborator names—are directly modified within the file that stores the blockchain, and any alterations are verified using the Blockchain Verifier tool, developed by Hyperledger Fabric. This test simulates a direct attack on the ledger's integrity, while the large scientific data files themselves remain untouched in their off-chain storage locations.

\section{Results}
\label{results}


    For integrity validation testing, the blockchain file was directly manipulated, which represents an extreme scenario in which the robustness of the data integrity mechanism can be demonstrated. By modifying the underlying ledger directly, the test bypassed application-level protections to challenge the blockchain’s core safeguards. The resulting hash mismatches clearly exposed the tampering, demonstrating how blockchain ensures tamper-evidence through its cryptographic chaining of transactions. Moreover, since each block is interlinked and verified through consensus, even isolated changes disrupt the ledger’s integrity and can be reliably flagged.
    
\begin{figure}[ht!]
        \centering
        \includegraphics[width=0.6\linewidth]{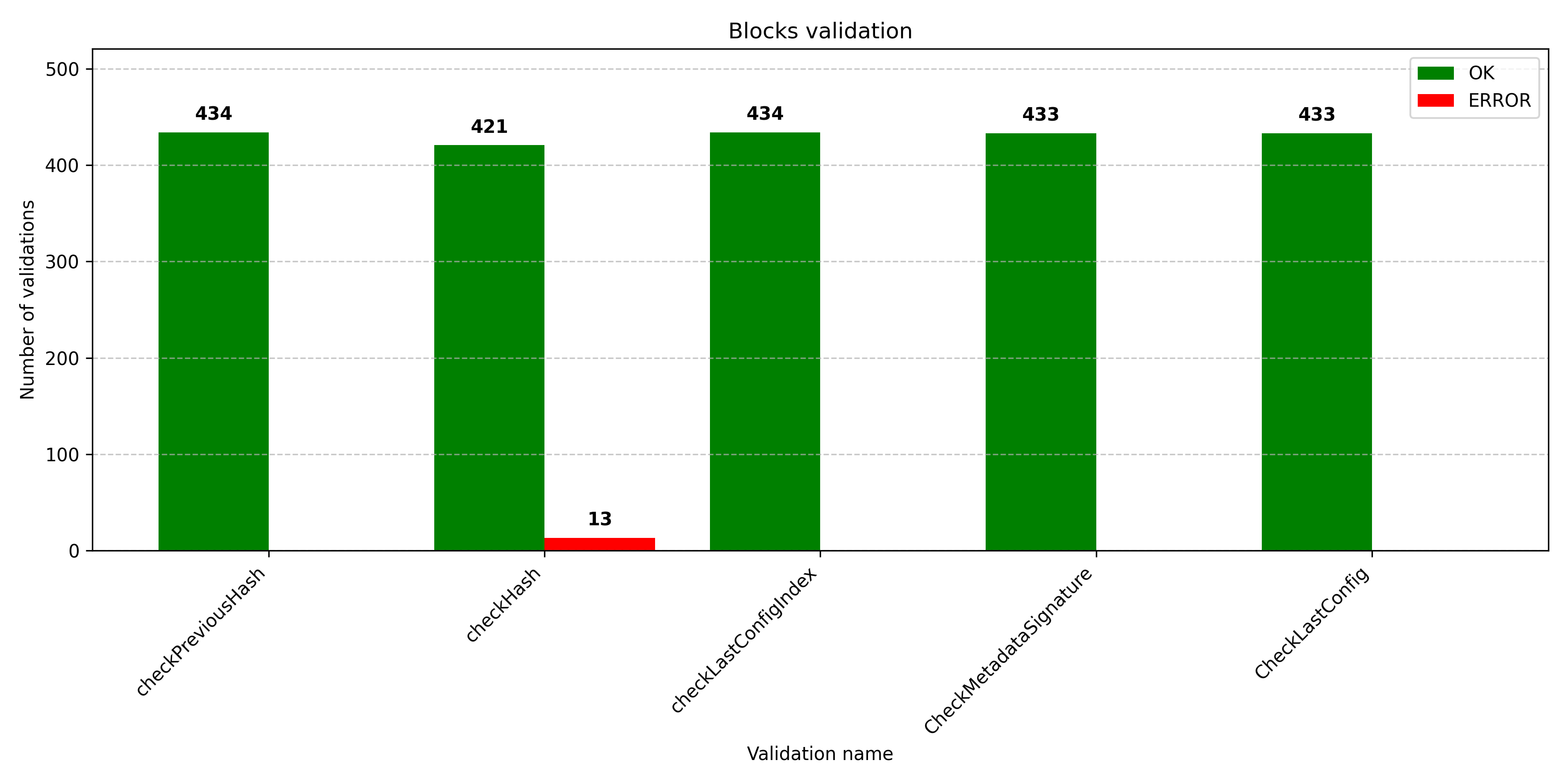} 
        \caption{Blocks validation}
        \label{fig:BlocksValidation}
\end{figure}

    However, this type of tampering is unlikely in a real environment. For this experiment, thirteen pieces of data were randomly selected from all transactions. The results of the block-level verification are shown in Fig. \ref{fig:BlocksValidation}, highlighting the hash mismatches that occur when on-chain data is tampered with.** The results are in format JSON \footnote{ Results repository: https://github.com/GSMier/ProyectoLago/tree/main/resultados }.

\begin{figure}[ht!]
        \centering
        \includegraphics[width=0.6\linewidth]{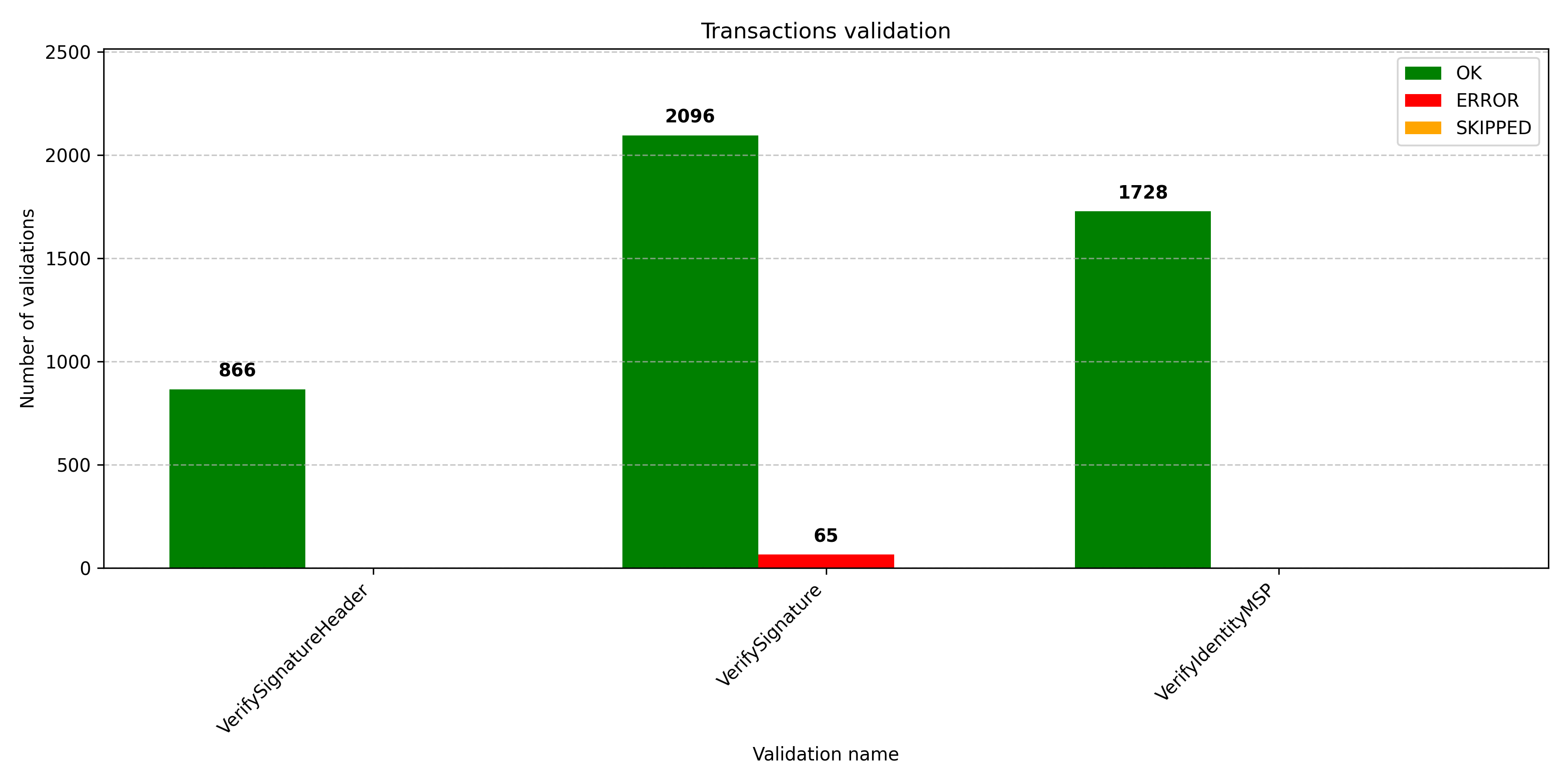} 
        \caption{Transactions validation}
        \label{fig:TxValidation}
\end{figure}

When analyzing the manipulated data with \textbf{Blockchain Verifier}, several specific anomalies were detected that confirmed the tampering. First, it was evident that the hash of the manipulated block's content no longer matched the original data hash stored in its header, which is the primary indicator of a data breach. The tool also revealed an unexpected behavior in Hyperledger Fabric: the hash of the block stored in its own header was not recalculated after the content was modified, which could complicate tampering detection if not for other checks. Furthermore, as shown in Figure \ref{fig:TxValidation}, the tampering invalidated numerous transaction signatures. In this case, \textbf{65 digital signatures} failed validation, confirming that the integrity of individual transactions was also compromised.

The validation of transactions in \textbf{Hyperledger Fabric} heavily relies on the digital signatures generated via the \textbf{Membership Service Provider (MSP)}. The MSP is responsible for issuing cryptographic identities to all authorized participants. Each transaction is then signed using the private key associated with a participant's identity. These digital signatures, which are directly dependent on the transaction's payload, serve two critical functions. First, they provide authentication, proving the transaction was created by a legitimate network member. Second, they ensure integrity, as any change to the transaction data after it has been signed would cause the signature verification to fail. This adds a crucial layer of security, as each transaction contains individually verifiable proof from its endorsers and owner, allowing any alteration to be quickly detected by a signature mismatch.

In this case, 65 signatures did not pass validation. This is because each transaction may contain one or more signatures, depending on the number of passing signatures required for validation. In the case of the \textbf{ScientificDataCollection} chaincode, four signatures are required to approve a transaction. It is important to consider that there are other chaincodes in execution, such as the one corresponding to the life cycle for the implementation of a new \textbf{chaincode}.

\section{Discussion}
\label{discussion}

Blockchain, as a decentralized ledger technology, offers a promising solution to various challenges faced by large-scale scientific infrastructures by enabling and supporting high-performance computing (HPC) distributed systems. Blockchain improves the use of dormant computational resources across the entire range of an advanced computing continuum, from edge devices to cloud computing capabilities and storage. Regarding applications, deploying blockchain to establish decentralized HPC marketplaces, in alignment with other blockchain initiatives, promotes the sustainability and shared economy of computational power\cite{Griffin2018}. However, within this context, the contribution of resources and data remains a contentious topic, requiring further discussion to assess the value of these resources and data accurately. For instance, in terms of cost efficiency, blockchain capitalizes on underutilized resources by allowing blockchain-based HPC platforms to significantly reduce the expenses related to large-scale computation compared to traditional supercomputers or cloud computing services. However, to ensure the benefits, it is crucial to enhance trust and security, and the blockchain's inherent transparency and immutability guarantee that computational tasks are carried out securely, data integrity supported and that participants are compensated.

More work must be done in order to prevent data manipulation. Due to the size of data samples for LAGO and many other HPC projects around the world, it is not feasible to store all data in the Blockchain, but recording hashes and other metadata must serve as a way for scientists to check data integrity and validate all research groups are working with the same samples independently of time and place.

\section{Conclusions and Further Work}
\label{Conclusion}

The development of a blockchain-based prototype for scientific data integrity demonstrated the feasibility of using Hyperledger Fabric to enhance secure access, traceability, and tamper-evidence in the management of large-scale distributed datasets processed in HPC clusters. Applied to a subset of the LAGO project’s public data, the prototype showed that blockchain can effectively support the immutability and verifiability of scientific records, even when stored in decentralized and heterogeneous environments.

The cryptographic mechanisms of the blockchain infrastructure ensure that any unauthorized alteration in the dataset is immediately detectable, fulfilling the requirements for scientific reproducibility and trust. By ensuring that the data has not been altered, it avoids the need to reprocess large volumes of information, a task that, in high-performance computing (HPC) contexts, involves a high energy cost. This benefit is particularly relevant in projects such as LAGO, where data is extensive and costly to generate.

During implementation, challenges arose in node synchronization and consistency, particularly when peers lost connectivity. This loss of connection directly impacted the prototype's reliability, creating a vulnerability where a node's local ledger could become inconsistent with the rest of the network. Submitting transactions to such a node resulted in errors due to the mismatch of data and hashes, threatening the system's integrity. These issues were addressed at the application level by implementing a mechanism to block transactions to peers in a corrupted state, while a re-initialization process was integrated to restore system consistency once connectivity was re-established. However, such challenges underscore the importance of further work in improving scalability, fault tolerance, and operational automation for deployment in full production environments.


Relying on other technologies in demand, such as Artificial Intelligence (AI), could benefit both in their weak points, in terms of traceability, interoperability, and scalability. Including AI  with blockchain and HPC is part of our ongoing work.

\begin{credits}
\subsubsection{\ackname} The authors thank the CYTED co-funded LAGO-INDICA network (524RT0159-LAGO-INDICA: Infraestructura digital de ciencia abierta) and the Universidad Industrial de Santander (UIS), especially the research group Cómputo Avanzado y a Gran Escala (CAGE), for their institutional support and the resources provided for this work. AI technology was used to proofread and polish this manuscript (OpenAI, 2025; Gemini, 2025).
\end{credits}


%
%
%
%

\end{document}